\documentclass{aa}

\usepackage[varg]{txfonts}
\usepackage[utf8]{inputenc}
\usepackage{float}
\usepackage{graphicx}
\usepackage{hyperref}
\usepackage{amssymb, graphics, setspace}
\usepackage{array}
\usepackage{pdfpages}
\usepackage{pdflscape}
\usepackage[center]{subfigure}
\usepackage{tikz}
\usepackage{siunitx}
\usepackage{natbib}
\begin{document}

\title{Survey of Orion Disks with ALMA (SODA)}
\subtitle{IV: Millimeter variability of protoplanetary disks}
\titlerunning{SODA IV. Millimeter-wavelength variability}

\author{D. Krebs\inst{1}
\and A. Hacar\inst{1}
\and J. Forbrich\inst{2, 3}
\and S. E. van Terwisga\inst{4}
}

\institute{Department of Astrophysics, University of Vienna, Türkenschanzstrasse 17, A-1180 Vienna, Austria
\and University of Hertfordshire, Centre for Astrophysics Research, College Lane, Hatfield AL10 9AB, UK \and Harvard-Smithsonian Center for Astrophysics, 60 Garden Street, Cambridge, MA 02138, USA \and Space Research Institute, Austrian Academy of Sciences, Schmiedl-Strasse 6, 8042 Graz, Austria 
}

\abstract{Short-term millimeter-wavelength variability in Class II young stellar objects (YSOs) has been considered minimal, often dismissed as very rare in ALMA-based studies on short timescales. However, recent findings suggest that variability on sub-day timescales is significant.}{This study aims to quantify short timescale ($\approx$1 hour to $\approx$3 days) millimeter flux variability in unresolved Class II YSO disks observed as part of the SODA survey, leveraging its large dataset to assess the statistical significance of such variability across the LDN 1641 and LDN 1647 regions in Orion.}{We analyze ALMA Band 6 observations of 873 YSOs obtained over multiple epochs using a custom flux extraction pipeline to separate individual execution blocks. By computing variability factors and statistical deviations from the mean flux, we identify sources exhibiting significant variability, accounting for noise and calibrator stability.}{Even with conservative selection criteria, we identify 20 variable sources within the sample ($\approx$5.3\% of all detections). The maximum of the measured variability factor is $\geq$7 over about one hour. More liberal selection criteria yield up to 60 sources that could be called potentially variable ($\approx$15.9\%).}{The number of variable sources detected in our sample indicates physical processes operating on time scales of hours, likely associated with magnetic reconnection and/or synchrotron radiation or strong cases of magnetic accretion changes. These detections show the need to reconsider typical stability assumptions in single-epoch observations and calls for follow-up studies in star-forming regions with ALMA. Masses derived from unresolved measurements could potentially be biased by this sort of variability.}
\date{Received date / Accepted date }
\maketitle
\nolinenumbers
\section{Introduction}

The study of protoplanetary disks and the dust they are made out of is instrumental for understanding processes like the formation of planets and whole planetary systems. Since its first light, disks have been observed with the Atacama Large Millimeter Array (ALMA) in the millimeter range, as the capabilities of it allow for a direct observation of the dust, both unresolved and resolved \citep{ALMA_2015}. As laid out in \cite{Manara2023} millimeter interferometry can estimate dust mass in protoplanetary disks as they trace thermal emission from dust in the outer regions of the disk and directly profit from the higher resolution capabilities of interferometry. Although uncertainties can arise from unclear optical depths and dust opacities, the resulting disk masses still offer valuable insights into the evolution of disks and formation of planets.

A typical dust temperature of 20~K and standard dust opacity are usually assumed to convert observed fluxes into dust mass estimates. Large surveys with ALMA have applied this method to disk populations in nearby star-forming regions, including the Lupus clouds \citep{Ansdell_2016, Alcala_2017, Ansdell_2018, Cieza_2019}, the 1 Myr Ophiuchus cloud \citep{Ansdell_2017} in $\sigma$Ori, \citep{Cieza_2019, Williams_2019}, and the older 5-10 Myr Upper Scorpius association \citep{Barenfeld_2016}. \citep{Pascucci_2016, Manara_2016} derive dust masses and relations to stellar masses in Cha I, \citep{Eisner_2018, van_Terwisga_2022} study the ONC and evaluate dust masses there, \citep{Cazzoletti_2019} in Corona Australis and \citep{Ruiz-Rodriguez_2018} in the IC 348 cluster \citep[see a full compilation in][]{Manara2023}. These surveys have achieved sensitivities down to sub-Earth mass levels, revealing that disk dust masses typically range from 1~M$_\oplus$ to $\approx$100~M$_\oplus$ , with a clear correlation between disk mass and stellar mass, and a systematic decline in average disk mass with age \citep{Ansdell_2018, Barenfeld_2016}.
Multiple surveys of YSOs have been carried out that found a change of flux over time, what we call variability from here on, i.e. the JCMT transient survey \citep[see][]{Mairs_2017, Johnstone_2022, Mairs_2024}, a survey of the Orion Nebula Cluster (ONC) \citep{Vargas_Gonzalez_2021} as well as several surveys that are not in the mm but in the cm using the VLA \citep{Forbrich_2016, Forbrich_2017, Marquez_2024}. 
Short timescale millimeter-wavelength variability has been observed in only a limited number of YSOs so far (mostly T Tauri stars) and including a number of binaries \citep{Bower_2003, Massi_2006, Salter_2008, Salter_2010, Mairs_2019, van_Terwisga_2020}. Additionally, flares outside of the mm have been known to occur for a longer time, see \cite{Garay_1987} and \cite{Forbrich_2008}, who find a variable cm source also in the ONC. Variability is here treated as the change of continuum flux over time in the mm continuum. Typically in literature, investigations of variability focus on typical timescales of weeks to many years which are thought to be related to a change in accretion of the surrounding material. In case of accretion driven variability, \cite{Fischer_2023} demonstrate that an increase of the period of the variability is correlated with an increase in its amplitude. However, other mechanisms can drive observable variable fluxes on different timescales. \cite{Benz_2010} note that (gyro-)synchrotron radiation can cause up to a factor of 30 of variability on < hour timescales, recently supported by the findings of \cite{van_Terwisga_2020} who find a source in NGC 2024 that changed by a factor of $\approx$30 over 24 hours, notably also in ALMA band 6 ($\approx$1.3~mm) , and \cite{Vargas_Gonzalez_2023} who find a source that varies by over a factor of 10 within mere minutes. \cite{Mairs_2019} note a source changing in flux by more than 50\% in 31 minutes. Recently, \cite{Miley_2025} have found a short term (50 min) variable source where a decrease in flux of over a factor 8 was detected. Some variability is also ascribed to binarity, see \cite{Kospal_2011, Massi_2006}. Also note \cite{Salter_2010}, who identify a binary system at $\lambda_{obs}$= 3~mm and interpret this as a consequence of reconnection between the magnetic fields of the binary companions at the periastron.
The total number of sources confirmed as variable at mm wavelengths in the literature is still low. \cite{Johnstone_2022} name only one among 1200 with the JCMT transient survey data. In \cite{Vargas_Gonzalez_2021} also one source out of around 130 is reported in a VLA catalog to be extremely variable, which in their case denotes any source whose flux density changes by a factor of 10 in an hour, after \cite{Forbrich_2017}. A later investigation in \cite{Vargas_Gonzalez_2023} found it to be variable over time scales of less than an hour by more than a factor of 30. 

The typically low rates of variable sources do not translate to other wavelengths however. \cite{Forbrich_2017} found 13 YSOs
to be variable in a sample of 507 with the VLA at 4~GHz-8~GHz (8~cm-4~cm) and the same 13 among a sample of 222 with Chandra in X-ray between 0.5~keV and 8.0~keV. These quite different findings show the need for a systematic investigation of variability in a statistically significant sample. Recently, \cite{Marquez_2024} have found more variability measured at cm wavelengths with the VLA over timescales of days to weeks, citing 10 out of 66 observed sources to be variable. Furthermore, they identified only one of those sources to have thermal origins, while the variability of the others was ascribed to non-thermal mechanisms, specifically magnetic reconnection. 

Quantifying continuum flux variability in the low mm range is important, as the mentioned surveys usually do not focus on shorter timescales below a few days to a week and in many cases, individual surveys use only one observation per source, making temporal variability harder to detect \citep[see][]{Miotello2022}.

We made use of the Survey of Orion Disks with ALMA (SODA)\footnote{\url{https://emerge.univie.ac.at/results/soda-survey/}} \citep[hereafter Paper I]{van_Terwisga_2022}. The SODA sample consists of 873 unresolved Class II young stellar objects (YSO) homogeneously surveyed with the Atacama Large Millimeter Array (ALMA). The SODA targets were selected from the full census of disky YSOs provided by previous infrared (IR) {\it Spitzer} surveys \citep{Megeath_2012,Megeath_2016} and include all Class II objects at least 4~pc south to the Orion Nebula Cluster (ONC) and across the LDN~1641 and LDN~1647 clouds. The SODA survey consists of multiple (2-3) observations for all sources observed over various timescales ranging from a day to three days. With this work, we aim to systematically analyze the variability of the SODA sample on short time scales, thus evaluating variability in a sample that was not intended for variability analysis.

Given the different original intent of the SODA data, we encourage the reader to consider this work a demonstration that a standard survey may exhibit a substantial number of variable sources, rather than a definitive work on variability.

\section{ALMA observations}
\subsection{Survey of Orion Disks with ALMA (SODA)}

The main goal of this project was to obtain a cloud-level demographics of the disk masses of Class II ($\sim$~2 Myr old) objects in Orion. SODA is the largest homogeneous proto-planetary disk sample obtained so far at millimeter wavelengths and almost doubles the number of ALMA observations of these types of objects in all nearby star-forming regions combined \citep{Manara2023}.

The entire SODA sample was surveyed using ALMA Band 6 observations (Project ID: 2019.1.01813.S; PI: van Terwisga) with the ALMA 12m array in its most compact configuration (C43-1) during November and December of 2019. These Band 6 observations included four broad-band 1.8~GHz wide spectral windows centered around 224~GHz, 226~GHz, 240~GHz and 242~GHz, respectively. Each SODA target, identified using a running [MGM2012] label \citep{Megeath_2012}, was observed using a single ALMA 12m pointing in snapshot mode (see Sect.~\ref{sec:datastructure}). The final maximum synthesised beam was found to be between $\theta_0\sim$~1.4-2.1\arcsec, which corresponds to $\approx$500-950~au at the distance of the LDN~1641 and LDN~1647 clouds of $\sim$454~pc \citep{Grossschedl_2018}.

Paper I extracted the total millimeter flux within a radius of 1.4\arcsec (or $\sim \theta_0$) from the field phase center. Mostly unresolved within the SODA beam, the authors converted the total integrated flux per source into equivalent total dust mass using standard procedures and dust properties, achieving an effective mass sensitivity of $\sim$1.5~M$_\oplus$ \citep[see][for a full discussion]{van_Terwisga_2022}. Different papers of this series used these SODA measurements to derive the distribution of Class II disk masses in different parts of the Orion A cloud (Paper I), as well as its dependence with the local UV-field \citep[][; Paper II]{vanTerwisga_2023_SODAII} and binarity \citep[][; Paper III]{Ricciardi_2025}. The extended disk population explored by SODA complements other similar surveys in more clustered regions in Orion such as the ONC carried out at both millimeter \citep{Mann_2014} and centimeter \citep{Forbrich_2016} wavelengths. 

\subsection{Data structure: access to the time domain information}\label{sec:datastructure}

\begin{table*}[ht]
\caption{SODA observations: Data structure}             
\label{tab:obs_SODA}     
\centering
    \begin{tabular}{c c c c c c} \hline \hline
        SB & Epochs (EBs) & Flux calibrator & Targets & MJD & Sampling\\
           &          &  & [MGM2012] & (58800+) & (hours)\\ \hline 
        1 & 3 & J0541-0541 & 1-175 & 8.27-9.30 & [$\tau_1$, $\tau_1$+1h, $\tau_1$+24h]\\
        2 & 3 & J0542-0913 & 176-361 & 11.28-14.23 & [$\tau_1$, $\tau_1$+1h, $\tau_1$+69h]\\
        3 & 2 & J0542-0913 & 362-574 & 14.27-14.35 & [$\tau_1$, $\tau_1$+1h]\\
        4 & 3 & J0542-0913 & 575-776 & 18.25-19.23 & [$\tau_1$, $\tau_1$+1h, $\tau_1$+24h]\\
        5 & 2 & J0542-0913 & 777-957 & 21.24-21.32 & [$\tau_1$, $\tau_1$+1h] \\
        6 & 2 & J0725-0054 & 958-1095 & 19.26-19.33 & [$\tau_1$, $\tau_1$+1h] \\ \hline
    \end{tabular}
\end{table*}

By construction, the standard operations of ALMA generate a particular data structure in observing projects including a large number of fields such as SODA \citep[see ALMA Handbook, ][for a full discussion]{ALMA_25}.
Targets within the same survey are grouped in individual Science Goals (SG) of no more than 150 ALMA fields which are observed as part of a single Scheduling Block (SB). Each SB may be carried out in different observational sessions, known as Execution Blocks (EB), limited to a maximum of total integration time of 1.5~hours per EB including both (flux, phase, and atmospheric) calibration and science observations. Each SB can be recursively observed in order to achieve the final sensitivity requested for each SG. In the particular case of SODA, the entire sample of 873 targets was divided in 6 SGs, each assigned to a single SB executed in 2 or 3 EBs depending on the SB and sometimes observed on different days (characterized by their modified Julian date, MJD). The integration time per (snapshot) field in each EB was approximately 16~sec, giving a total integration time per source of $\sim$30-45~sec. We list the main properties of our ALMA observations in Table~\ref{tab:obs_SODA}.

Disk mass measurements in the (sub-)mm continuum rely on the consistent measurement of fluxes. Similar to previous ALMA works \citep[e.g.,][]{Miotello_2017}, Paper I concatenated all individual EBs available per target in order to maximize the sensitivity of their maps. Our new Paper IV uses a different strategy in order to investigate the millimeter flux variations over time. By reducing each EB separately (see Sect.~\ref{sec:datareduction}) we obtained flux measurements of 2-3 individual epochs per source. 

Several caveats should be considered when investigating these new flux measurements. SODA was not designed to carry out time domain studies. Compared to other systematic studies targeting specific time samplings \citep[e.g.,][]{Vargas_Gonzalez_2021}, the time between EBs was arbitrarily determined by the ALMA scheduler and queuing priorities. The starting time for each observation ($\tau_1$) and the time between consecutive EBs ($\Delta \tau$) were thus randomized and may be different for each SB. Likewise, the number of EBs or epochs varies between SBs depending on the number of executions (2 or 3) needed to achieve the target sensitivity given the atmospheric and observing conditions on each day (see Table~\ref{tab:obs_SODA}).

Our new study leverages the statistically significant sample size of SODA to study the frequency of mm wavelength variability in a population of protoplanetary discs. All targets within each SB ($\sim$~150) are visited with the same frequency including regular observations of the same flux calibrator (see also Sect.~\ref{sec:PhaseCalib}). Luckily, all SB were also observed with two consecutive EBs providing a uniform data sampling of $\Delta \tau\sim$~1~h across our entire sample. In three cases, additional EBs allow to also explore flux variations on the order of $\Delta \tau\sim$~24~h (SBs \#1 and \#4) and $\Delta \tau\sim$~69~h (SB~\#2), respectively (see Table~\ref{tab:obs_SODA}).
This randomized data structure provides a unique unbiased dataset to explore time variability at millimeter wavelengths, particularly in timescales of $\Delta \tau\sim$~1~h (all SBs).
We note that this same analysis strategy might be also relevant in future time analysis and archival research with ALMA to assess flux variability in other surveys not necessarily designed for these purposes.

\subsection{Data reduction}\label{sec:datareduction}

For all our new reductions we make use of Common Astronomy Software Application \citep[CASA;][]{CASA_2022} as standard data processing software for ALMA observations. Data calibration was carried out in a similar way as in \cite{van_Terwisga_2022}. We extracted all individual visibilities per ALMA source from the corresponding raw visibility file and concatenated into a single dataset. Each map is then individually de-convolved using \texttt{tclean} task in \texttt{mfs} continuum mode using a standard Briggs weighting with \texttt{robust} parameter equal to 0.5, 10\,000 iterations and a threshold of 0.1~mJy imaged within an area of $400\times400$~pix$^2$ with an individual pixel size of d$_{pix}=0.08$\arcsec. Different tests show variations of less than 10\% in flux when using different deconvolution parameters (e.g., pixel size of d$_{pix}=0.25$\arcsec). In addition, we performed tests with a different number of iterations, Briggs weighting, and varying cleaning threshold showing no significant differences, at worst measuring 15\% of individual fluxes, while the average across the full sample only changed by $<$5\%, which demonstrates the robustness of the process against our specific choice of parameters. We emphasize that the aim of this work is to demonstrate the presence and approximate scale of short term variability in an otherwise standard survey data set, not to deliver a conclusive census of variable sources. As shown above, our results are robust to reasonable choices in reduction parameters. The exact count crossing any given threshold remains more sensitive to methodological choices near the low-variability end of the distribution, as expected for a hard cut applied to a quantity with uncertainty.

After imaging, we convolved each map using the task \texttt{imsmooth} into a final Gaussian circular beam $\theta_{conv}$ similar to the largest synthesized beam $\theta_0$ found in all EB in each field. These final images have $\theta_{conv}$ values ranging between $\sim$~1.4\arcsec and 2.1\arcsec. 
This is usually larger than the original beam size in Paper I (i.e. $\sim$~1.4\arcsec). By adopting a similar beam in all epochs we aim to homogeneously analyze the flux variations per source.

Our data reduction methodology follows the massive processing strategy implemented by \citet{van_Terwisga_2022} in Paper I, this time also applied to individual EBs. 
We carried out all our reductions in the {\it Vienna Scientific Cluster 5} (VSC-5)\footnote{Recently re-branded as Austrian Science Computing: \url{https://asc.ac.at/}} (project ID: 71908) which allows the parallelization of the CASA deconvolution process using multiple and simultaneous threads \citep[see also][for a similar data reduction scheme]{Hacar_2024}.
Using CASA version 6.5, firstly we constructed a single concatenated ALMA image per target (873 images) combining all EBs available per field. Secondly, we carried out individual deconvolutions per EB (or epoch) obtaining 2 or 3 maps per target (1946 snapshots in total). Each image is cleaned individually using CASA in parallel mode run on VSC-5. The large number of available nodes in the VSC-5 supercomputer allows to process the entire set of close to 2000 images in our sample in roughly 6 hours.

\section{Characterizing flux variations}

\subsection{Data overview: observed flux variations}

\begin{figure*}[ht!]
    \centering
    \includegraphics[width=0.77\textwidth]{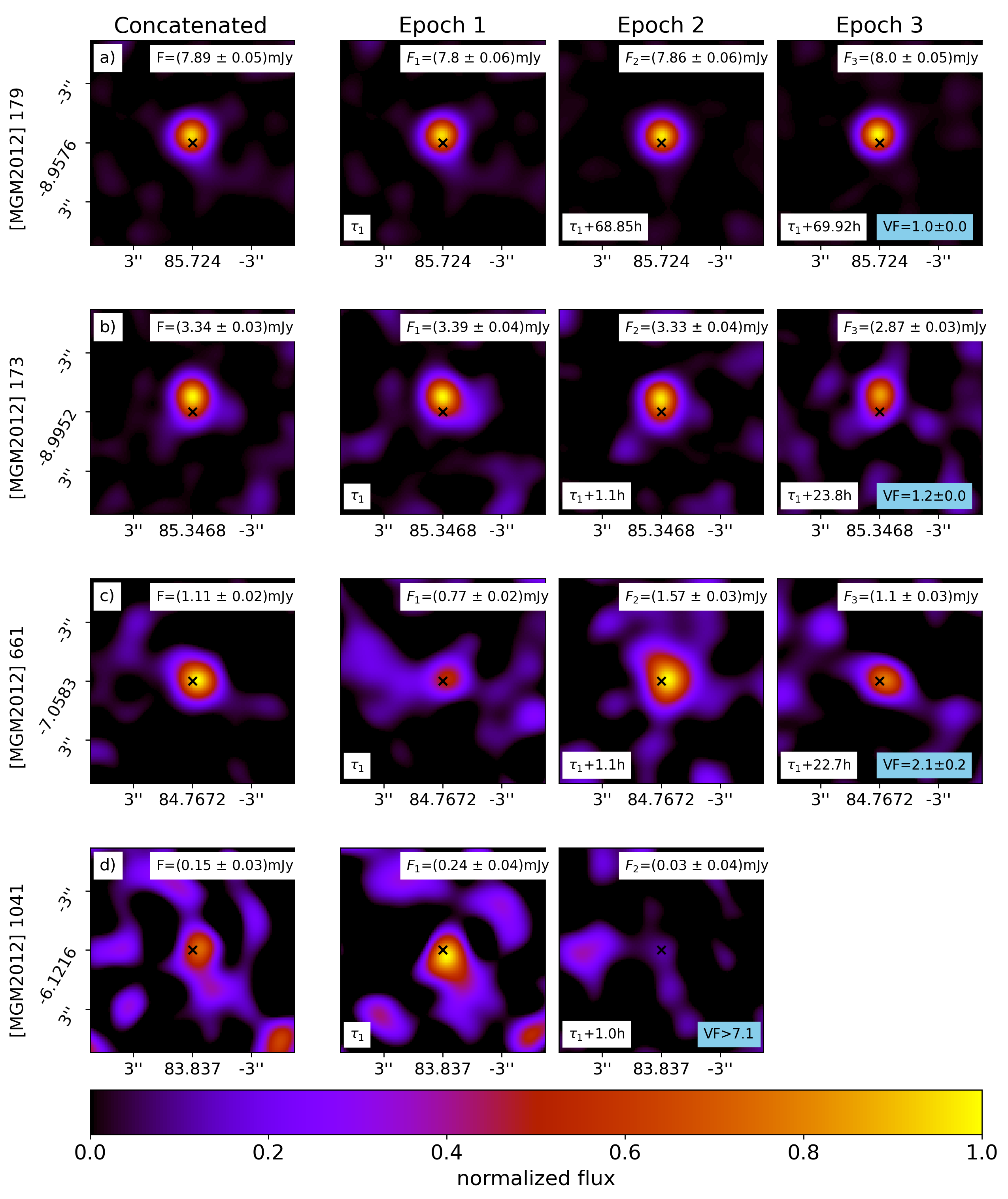}
    \caption{Flux variations in four representative targets of our SODA sample (rows) including both their concatenated ALMA map (first column) as well as the individual maps per epoch (second to fourth columns).
    Sources are sorted on increasing variability factor f$_{var}$ (Sect.~\ref{sec:VF}) of each time series (bottom right corner in the rightmost panel). From top to bottom: {\bf (a)} [MGM2012]~179 (no variability), 
    {\bf (b)} [MGM2012]~173 and  {\bf (c)} [MGM2012]~661 (intermediate variability), 
    and {\bf (d)} [MGM2012]~1041 (high variability). 
    In each row the image fluxes are normalized with respect to the maximum peak flux value of the series (see color bar).
    Each sub-panel includes the time difference ($\Delta \tau$) with respect to the first observation $\tau_0$ (top left corner) and the corresponding total flux $F$ (top right corner).
    The black crosses in the center of the image indicate the nominal Spitzer positions \citep{Megeath_2012}. There is no significant offset between the nominal Spitzer positions and the perceived source position in this work.}
    \label{fig:dataoverview}
\end{figure*}

We display a series of four SODA targets representative of our observations in Figure~\ref{fig:dataoverview}. In each row we show the individual continuum maps both concatenated and per epoch (2 or 3 depending on the SB where the source was observed) in different columns, using normalized flux units with respect to the brightest peak of the series. The phase-center of all images corresponds to the position of the targets identified by {\it Spitzer} (black cross). Offsets between our continuum detections are usually within one $\theta_{conv}$ with respect to the image phase-center and below the {\it Spitzer} resolution \citep[$\sim$~2-5\arcsec;)][]{Megeath_2012}. As discussed in Paper I, most of our continuum sources ($>$95\%) remain unresolved. 

To illustrate the properties of our sample, we ordered our targets (from top to bottom) by the level of variations observed in their integrated flux (see flux values on the top right corner of each sub-panel; Sect.~\ref{sec:fluxextracton}).  
Most SODA sources (for example [MGM2012]~179, panel~a) show roughly constant flux values with no significant (or only minor) flux variations across epochs. On the other hand, some bright sources show relative variations of $\sim$~20\% (for example [MGM2012]~173, panel~b). In exceptional cases, these flux changes can be up to more than a factor two (>200\%) (for example [MGM2012]~661, panel~c). In some other cases, similar flux variations can make the source (un-)detected in our sample depending on the epoch (for example [MGM2012]~1041, panel~d). 
Despite our relatively shallow coverage, these observations demonstrate the potential of our SODA sample to detect flux variability at millimeter wavelengths.

\subsection{Flux measurements}\label{sec:fluxextracton}

The systematic analysis of the entire SODA observations requires the homogenization and automation of the flux extraction method for data observed in different epochs and with different sensitivities.
We first characterize the source position in each of our SODA targets from the analysis of the concatenated image combining all epochs. There we identify the position of the source after looking for the brightest pixel within a radius of 20 pixels (or $\sim \theta_{conv}$) from the image phase-center in order to prevent confusion  in the case of close binaries (see also Paper III).  
Starting from this central position, we identified those adjacent pixels with fluxes within a contour defined by four times the standard deviation $\sigma_{RMS}$ of the whole image (both in units of Jy~beam$^{-1}$). We then add the fluxes of all $N_{clip}$ selected pixels to estimate the total source flux 
\begin{equation}\label{eq:flux}
    F=\sum_{pix=1}^{N_{clip}} \frac{f_{pix}}{pix/beam}
\end{equation}
where the number of pixels per beam $pix/beam=\frac{A_{beam}}{A_{pix}}$ is given by the ratio between the corresponding pixel $A_{pix}=d_{pix}^2$ and beam $A_{beam}=\frac{\pi}{4}\theta_{conv}^2$ areas, respectively.
On the other hand, we estimate an upper limit of the corresponding flux uncertainty within a similar number of pixels as $\sigma = \sqrt{N_{clip}}\cdot \frac{\sigma_{RMS}}{N_{pixbeam}}$ which can be approximated as
\begin{equation}\label{eq:sigmaflux}
    \sigma\sim\frac{\sigma_{RMS}}{\sqrt{N_{clip}}}.
\end{equation}
if the source fills the beam, that is $N_{clip}\sim pix/beam$. 

Note that for the uncertainty calculation, 10 pixels around the area where F was determined were not taken into account for $\sigma$ since they are correlated with F and would artificially increase the noise based on the brightness of the source.

Following Eqs.~\ref{eq:flux} and \ref{eq:sigmaflux} we determine the flux values in the concatenated image F$_0\pm \sigma_0$. We then use those pixels identified in this high sensitivity image as mask in all EBs and obtain similar flux estimates for each individual epoch F$_i\pm \sigma_i$ (with $i=1,2,3$) using the same formulae (e.g., Fig.~\ref{fig:dataoverview}~c). As additional advantage, the above pixel mask allows to derive flux upper limits in weak sources only detected in the concatenated and/or single epochs (see Fig.~\ref{fig:dataoverview}~d). 

Our flux extraction method differs from other more conventional approaches. 
The total flux of unresolved sources such as those in our SODA disks could be extracted from the peak flux value in our ALMA maps (expressed in units of Jy~beam$^{-1}$). For instance, \citet{van_Terwisga_2022} used this method in order to retrieve the flux measurements in the original Paper I. 
Instead, our flux extraction method aims to minimize the effects of noisy pixels and the spurious identification of false positives. Although it is the case in our sample \citep[see][]{van_Terwisga_2022}, our method does not require sources to be strictly unresolved.
By comparing these two approaches, we find a linear correlation between our F$_0$ flux estimates with those obtained in Paper I. Yet, our values are systematically underestimated by a factor of 9\% on average. 
These differences are however expected by the $4\cdot\sigma_{RMS}$ clipping applied to our images compared to the full Gaussian profile considered by van Terwisga et al. We choose a conservative $4\cdot\sigma_{RMS}$ to statistically have no pixel above the cutoff in the beam. As a result, and compared to the 502 sources reported in Paper I, we detect 377 sources with S/N$\ge$~4 (or 75\%) with our method, where most of these discrepancies are attributed to weak detections in the original SODA sample. 

\subsection{Alternative flux extraction}
\label{sec:alt_flux}

A comparison with more conventional flux extraction methods as e.g. in \cite{van_Terwisga_2022}, who measure integrated flux from the brightest pixel value in each observation approximating an unresolved gaussian disk, has been carried out. If flux is extracted this way, four more variable candidates make the 1.2 VF threshold. We must emphasize that this may reasonably be expected from a sharp cutoff and a certain distribution of VFs, all of which have an uncertainty, with any given method.

If we additionally reduce the data with \texttt{robust} of 2 and without applying smoothing to images, 27 total variable candidates can be found, while our original method produced 20, 12 of which are in the 27. Note that the top 60\% of variable candidates (i.e. the top 12 from the method described herein) can be found independently of the approach or detailed method. The exact count is sensitive to where the detection threshold is placed: our variability factor (VF) carries an uncertainty, while the threshold cut itself does not, so sources near the boundary can reasonably shift across it under small changes in the reduction. This is an expected consequence of applying any hard cut to a quantity with error bars. Given that we aim to be conservative throughout, this movement is expected to affect only borderline cases and does not bias the overall result.

\section{Millimeter variability in the SODA sample}

Following the results shown in Fig.~\ref{fig:dataoverview}, the main goal of this Paper IV is to quantify the flux variations across the SODA sample. Note that this should not be understood in an absolute sense, but more as an experiment into what variability may go undetected in a typical survey of this kind. We do not claim to know the exact number of variable sources and rather see them as variable candidates in the sample that could lead observers to inaccurate conclusions if their variability is ignored.
Investigating flux variability in our survey presents some challenges, however.
On one hand, the limited number of epochs available (2 or 3 per source) prevents the use of some of the advanced statistical analyses applied to the more densely sampled light curves observed in other dedicated millimeter surveys \citep[e.g.,][]{Mairs_2017,Vargas_Gonzalez_2023}. On the other, the non-negligible likelihood of false positives in noisy observations of unresolved sources requires a careful estimate of the individual image noise level and requires the use of conservative criteria when identifying variable sources. 

\subsection{Stability and intrinsic flux fluctuations}\label{sec:PhaseCalib}

\begin{figure}[h]
    \centering
    \includegraphics[width=\linewidth]{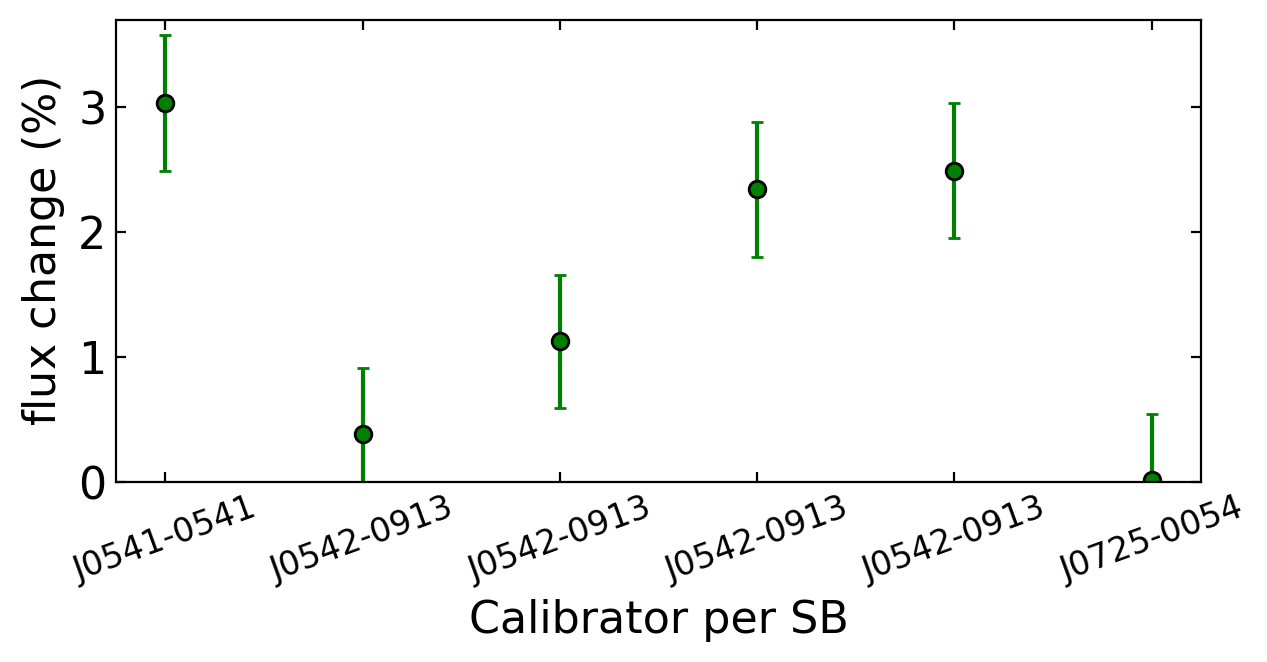}
    \caption{Flux variations (mean and standard deviation; green circles plus error bars) in the observed flux calibrators per SB (see Table~\ref{tab:obs_SODA}).}
    \label{fig:phasecal}
\end{figure}

We display the observed flux changes (in \%) within the different flux calibrators per SB in Fig.~\ref{fig:phasecal} (see also Table~\ref{tab:obs_SODA}). 
The flux values observed in our flux calibrators (see table \ref{tab:obs_SODA}) are observed to vary in $\lesssim3$\% comparing different observing sessions and SB.
These low flux variations obtained in our ALMA 12m observations are comparable to the flux uncertainties obtained in the dedicated ACA observations by \citet{Francis_2020}.

\subsection{Variability factor}\label{sec:VF}

\begin{figure}[ht]
    \centering
    \includegraphics[width=1\linewidth]{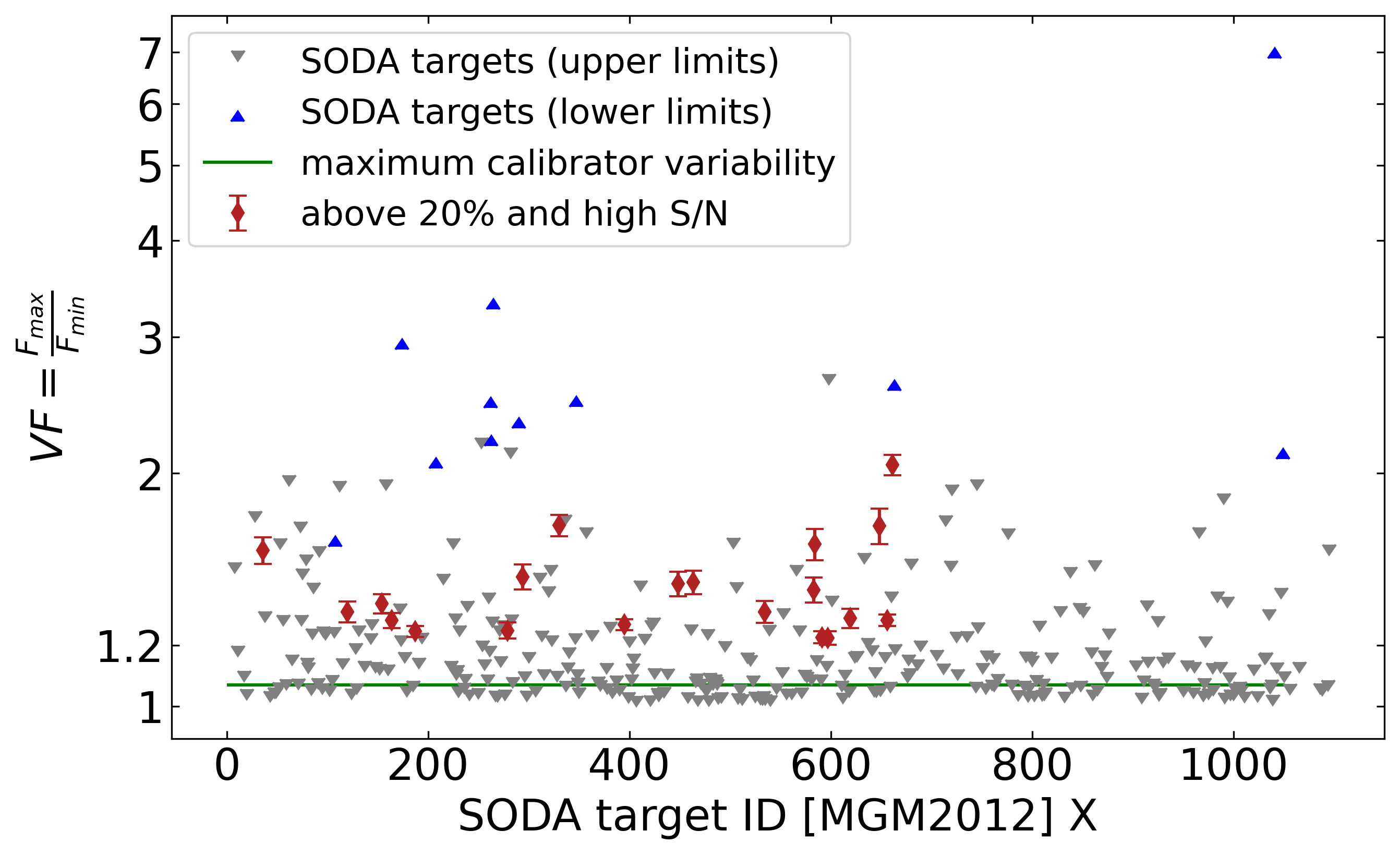}
    \caption{Variability factor ($VF$) for all the SODA sources with fluxes higher than 4$\sigma_i$ identified by their [MGM2012] catalogue ID. Different symbols identify distinct detections in our survey: (grey inverted triangles) upper limits, (blue triangles) lower limits, and (red diamonds) variable sources (i.e. $VF>1.2$ and $\frac{VF}{s_{VF}}>7$; see text for a discussion).
    }
    \label{fig:VF}
\end{figure}

In the following sections we make use of two independent variability metrics considering the characteristics of our science sample (377 targets).
First, we use the variability factor \citep[VF,][]{Vargas_Gonzalez_2021} defined as the ratio between the maximum (F$_{max}$) and minimum (F$_{min}$) flux values in all epochs observed, that is
\begin{equation}\label{eq:VF}
    VF= \frac{F_{max}}{F_{min}}.
\end{equation}
Estimates of VF highlight the (relative) maximum flux variation across epochs in individual targets. As principal advantage, errors in these VF estimates $s_{VF}$ are easily derived from the direct propagation of the flux errors in Eq.~\ref{eq:VF}.

We show the estimates on $VF$ and $s_{VF}$ obtained in the 377 targets selected in our sample in Fig.~\ref{fig:VF} ordered by the source ID. In most cases, we display an upper limit of the corresponding $VF$, i.e. $VF+4\cdot s_{VF}$. We use $4 s_{VF}$, i.e. four times the standard deviation of the specific variability factor, as with the large number of sources, we can be confident that none are likely to be outliers. Only in cases of epochs with no-detection (e.g., Fig.~\ref{fig:dataoverview}, panel d) we report the $VF$ values as lower limits assuming F$_{min}=\sigma_i$. As expected, most sources show roughly constant fluxes characterized by low variability factor values around $VF\sim 1$. For comparison, the variability factor of our flux calibrators is typically of $VF\sim1.03$ (see Sect.~\ref{sec:PhaseCalib}). Note that sources that are only detected in some epochs are treated as upper limits with no distinction of how many epochs show the source.

Several sources in our sample show values of $VF>1$, although their variations might not be significant. Noisy detections and/or low-intensity sources are prone to show higher $VF$ values. 
To test the quality of our method we created a distribution with the same spread as our sources described by the $VF$ values with mean values of 1 and standard deviation of 0.03 (similar to our flux calibrators) from which we draw (bootstrap) three random values. This is to test our method with statistical data to rule out outliers. We obtain a maximum $VF\sim1.15$ considering this synthetic sample, with no single source above $VF>1.20$. Detections beyond this $VF=1.2$ (=20\%) value are therefore unlikely to be caused by random noise.

Based on the results above we conservatively identify as variable only those sources that simultaneously satisfy the following two conditions: (a) their flux varies more than a 20\%, that is $VF>1.2$, and (b) that this variation is significant, for which we define $\frac{{VF}}{ s_{VF}}>7$ as threshold. Using these criteria we identify a total of 20 variable candidates within our 377 targets (or $\sim$~5.3~\%). Among them, source [MGM2012]~661 appears as the source with higher variability showing a factor of two change in flux ($VF\sim2$) within the 3 epochs surveyed by SODA (see also Fig.~\ref{fig:dataoverview}, panel c). 

The lack of variable candidates from SODA target 670 onward may be explained by a significant enlargement in the average source diameter $\gtrapprox$2\arcsec, with earlier sources being usually $\lessapprox$1.5\arcsec.

The threshold was set to approximate the highest variability plausibly attributable to calibrator uncertainty when expressed in $\frac{{VF}}{s_{VF}}$, and was kept fixed throughout this work. This was assessed by bootstrapping data with a standard deviation of 3.2 percent and simulating $10^3$ data points, the rough number of sources in the SODA sample. At $\frac{{VF}}{s_{VF}}$ $>$ 4, no simulated non-variable source was above the cutting line. $\frac{{VF}}{s_{VF}}$ $>$ 7 was chosen as a conservative threshold should the calibrator uncertainty not be the only systematic source of uncertainty. 
Other studies have adopted a conservative maximum calibrator uncertainty of 10\% for ALMA Band 6 observations (\citep{Ansdell_2018, Yoshida_2025}; our VF = 1.2 threshold can similarly be read as a worst-case upper limit on variability set by calibration.

\begin{table}[h]
    \centering
    \caption{The number of variable candidates dependent on the $\frac{{VF}}{s_{VF}}$ threshold}  
    \begin{tabular}{c|ccccc}
        $\frac{{VF}}{s_{VF}}$ & 4 & 5 & 6 & 7 & 8\\ \hline
        \# variable candidates& 47 & 36 & 28 & 20 & 15\\
    \end{tabular}
    \label{tab:VF cutoffs}
\end{table}

Table \ref{tab:VF cutoffs} shows the influence of different $\frac{{VF}}{s_{VF}}$ thresholds on the number of variable candidates. Note that at a threshold of below 4, the number of candidates is significantly higher, partially including very faint sources and sources that potentially vary by only 5-10 \% and could be unjustly amplified by calibrator uncertainties. In this work, we have chosen to adopt $\frac{{VF}}{s_{VF}}>$ 7 as our conservative selection criterion in order to preserve only the variability candidates that were detected with high confidence.

\subsection{Statistical variability}

\begin{figure}[ht]
    \centering
    \includegraphics[width=\linewidth]{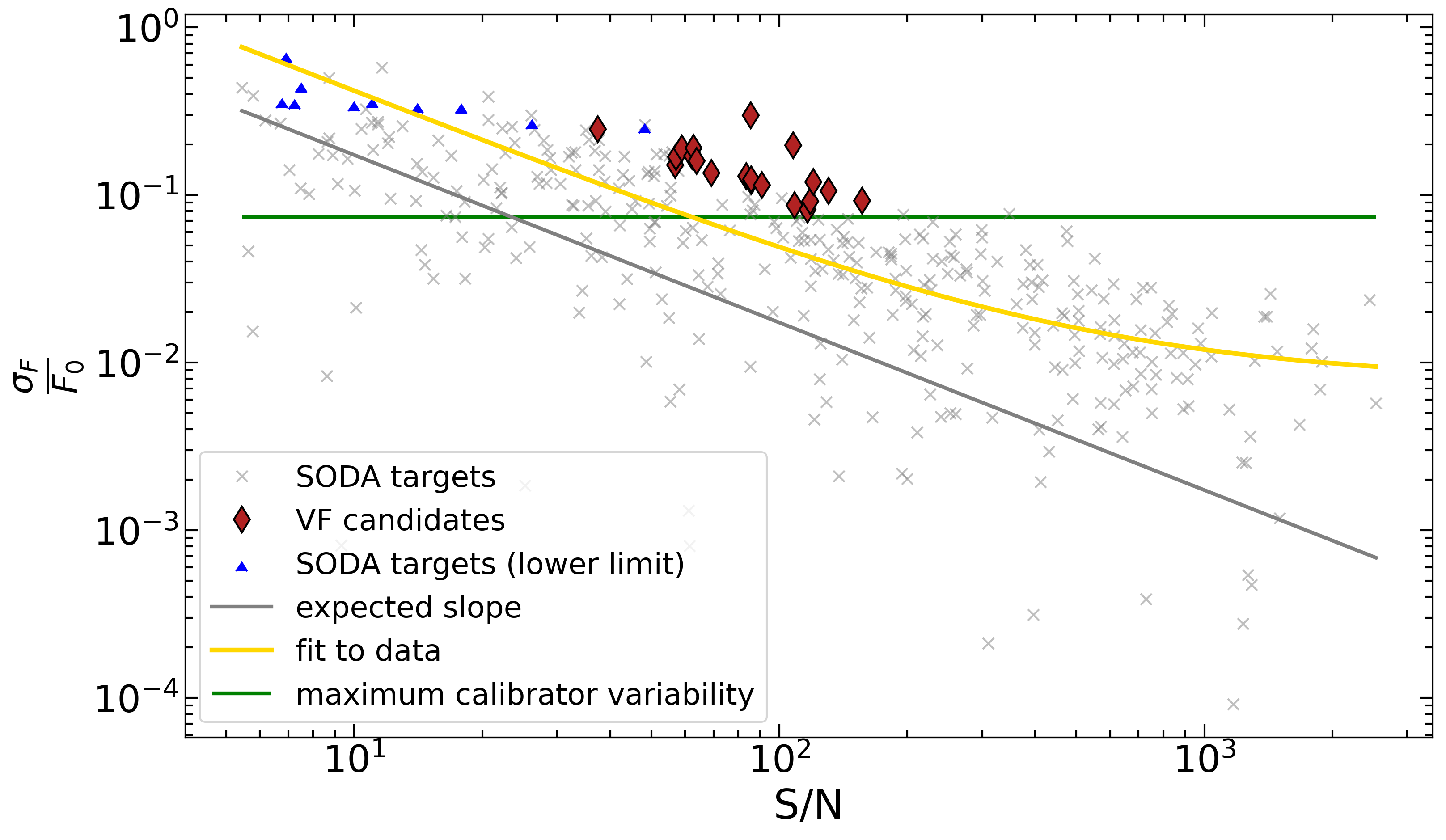}
    \caption{Normalized flux variation ($\frac{\sigma_F}{F_0}$) for all SODA sources with fluxes higher than 4$\sigma_i$ vs. the signal to noise ratio S/N of the concatenated image for all sources. The symbols identify distinct detections in the survey: (grey crosses) SODA targets, (blue triangles) lower limits and (red diamonds) variable sources (i.e. $VF>1.2$ and $\frac{VF}{s_{VF}}>7$). The yellow line denotes the fit to the whole data set with the expected relationship as a gray line. The green line shows the maximum calibration variability. The grey crosses above the fit are not treated as variable since they do not meet our criteria for VF due to large uncertainties. Note that maximum calibration variability most pessimistically leads to a $\frac{\sigma_F}{F_0}$ of 7.4~\% and a VF of 6.6~\%.
    }
    \label{fig:VarSN}
\end{figure}

Second, we also characterized the flux variations in our targets using the standard deviation considering the individual F$_i$ and mean $\Bar{F}$ flux measurements in all observed $N$ epochs such as
\begin{equation}\label{eq:sigmaF}
    \sigma_F = \sqrt{\frac{1}{N}\sum_i^N \left(F_i - \bar{F} \right)}.
\end{equation}
We estimate our $\sigma_F$ errors by including Gaussian flux fluctuations in each F$_i$ measurement given its individual $\sigma_i$.  
Overall, $\sigma_F$ should statistically approach the true flux $\sigma$ in case of pure noise variations. 
Given our limited number of epochs ($N\le3$) we notice that the value of $\sigma_F$ can however fluctuate and large deviations from the true standard deviation ($\sim \sigma_i$) are expected when investigating individual sources. As demonstrated below, deviations from these expectations can be statistically evaluated when combining the analysis of the entire SODA sample. 

Pure noise fluctuations $\sigma_F$ are expected to have lower relative impact with respect to their mean flux F$_0$ in bright sources detected at high signal-to-noise ratios ($S/N$) since
\begin{equation}\label{eq:VarNoise}
    \frac{\Delta F}{F_0}=\frac{|F_i-F_j|}{F_0}\sim\frac{\sigma_F}{F_0}\simeq\frac{1}{S/N}.
\end{equation}
Eq.~\ref{eq:VarNoise} depicts the intrinsic anti-correlation between the normalized flux variations $\frac{\sigma_F}{F_0}$ and the source $S/N$ given by the noise of the data.
We show the observed $\frac{\sigma_F}{F_0}$ values as function of $S/N$ (including error bars) in Fig.~\ref{fig:VarSN} both displayed in logarithmic scales where the dependence
\begin{equation}\label{eq:VarNoise_log}
{\rm log}\left(\frac{\sigma_F}{F_0}\right)=-{\rm log}(S/N)
\end{equation}
is expected. Overall, the bulk of our sources follow the above anti correlation for the typical noise of our data, i.e. $<\sigma_F>=0.05$~mJy (grey line; see also Sect.~\ref{sec:fluxextracton}). A fit to all our data points (in log-log space), retrieves a similar result to these predicted dependence (yellow line) with essentially the same slope as the expected relation for lower S/N ratios. Note the expected relation goes to very low $\frac{\sigma_F}{F_0}$, where in reality they reach a seemingly stable minimum of roughly 1\% $\frac{\sigma_F}{F_0}$. 

Similar to figure \ref{fig:VF}, the red diamonds in Figure \ref{fig:VarSN} represent the VF candidates. They lie in an area above the maximum calibrator variability (green line) and as far away from the fit of the data as can be in our data. Considering our conservative approach, these 20 sources make up 2.3~\% of our sample and 5.3~\% of all detections with source [MGM2012]~661 as the most prominent example of this variation. 

We modified Eq.~\ref{eq:VarNoise_log} to quantify larger deviations from this trend such as
\begin{equation}\label{eq:VarNoise_logerr}
{\rm log}\left(\frac{a<\sigma_F>}{F_0}\right)={\rm log}\left(\frac{<\sigma_F>}{F_0}\right)+{\rm log}(a)=-{\rm log}(S/N),
\end{equation}
where $a$ denotes the relative flux variation (in $\sigma$ units)  of a source  with respect to the flux fluctuations expected due to pure noise and where $a$ may be understood as an indication of how far away from a theoretical prediction a specific source lies.

\begin{table}[ht]
\caption{VF candidates and their statistical variability}  
\label{tab:VF}     
\centering
    \begin{tabular}{c c c} \hline \hline
        [MGM2012] & VF & $\frac{\sigma_F}{F_0}$ (\%) \\ \hline
        36 & 1.59 & 19.1 \\
        120 & 1.32 & 12.1 \\
        154 & 1.36 & 12.9 \\
        164 & 1.29 & 12.4 \\
        187 & 1.25 & 9.2 \\
        279 & 1.25 & 8.7 \\
        294 & 1.47 & 15.1 \\
        330 & 1.71 & 19.8 \\
        395 & 1.28 & 12.0 \\
        448 & 1.44 & 17.1 \\
        463 & 1.45 & 16.9 \\
        534 & 1.32 & 13.5 \\
        583 & 1.41 & 16.0 \\
        584 & 1.62 & 19.0 \\
        591 & 1.23 & 9.2 \\
        597 & 1.23 & 8.2 \\
        619 & 1.30 & 11.5 \\
        648 & 1.71 & 24.7 \\
        656 & 1.29 & 10.6 \\
        661 & 2.05 & 29.8 \\
    \end{tabular}
\end{table}

Table \ref{tab:VF} shows the VF candidates (red diamonds) identified in figures \ref{fig:VF} and \ref{fig:VarSN} with their corresponding variability factor VF and the standard deviation $\frac{\sigma_F}{F_0}$ in percent. We would like to emphasize that  the standard deviations $\frac{\sigma_F}{F_0}$ of all these sources are higher than what can be expected from the most pessimistic calibrator variability. It should also be emphasized that 12 of the VF candidates reach $a\geq5$, while all 20 VF candidates are $a\geq3$. This shows that the variability is substantially stronger than what would immediately be expected from a simple theoretical consideration.

\subsection{Variability timescales}\label{sec:timescales}
Lastly, we investigated the timescales over which the variability occurs. Depending on the particular SB, (see table \ref{tab:obs_SODA}) observations allow us to detect variability on scales of 1 hour, 1 day and 3 days. Since there are more observations over the short timescales, as expected most of the variabilities fall into the 1 hour timescale.  

\begin{figure}[ht!]
    \centering
    \includegraphics[width=1\linewidth]{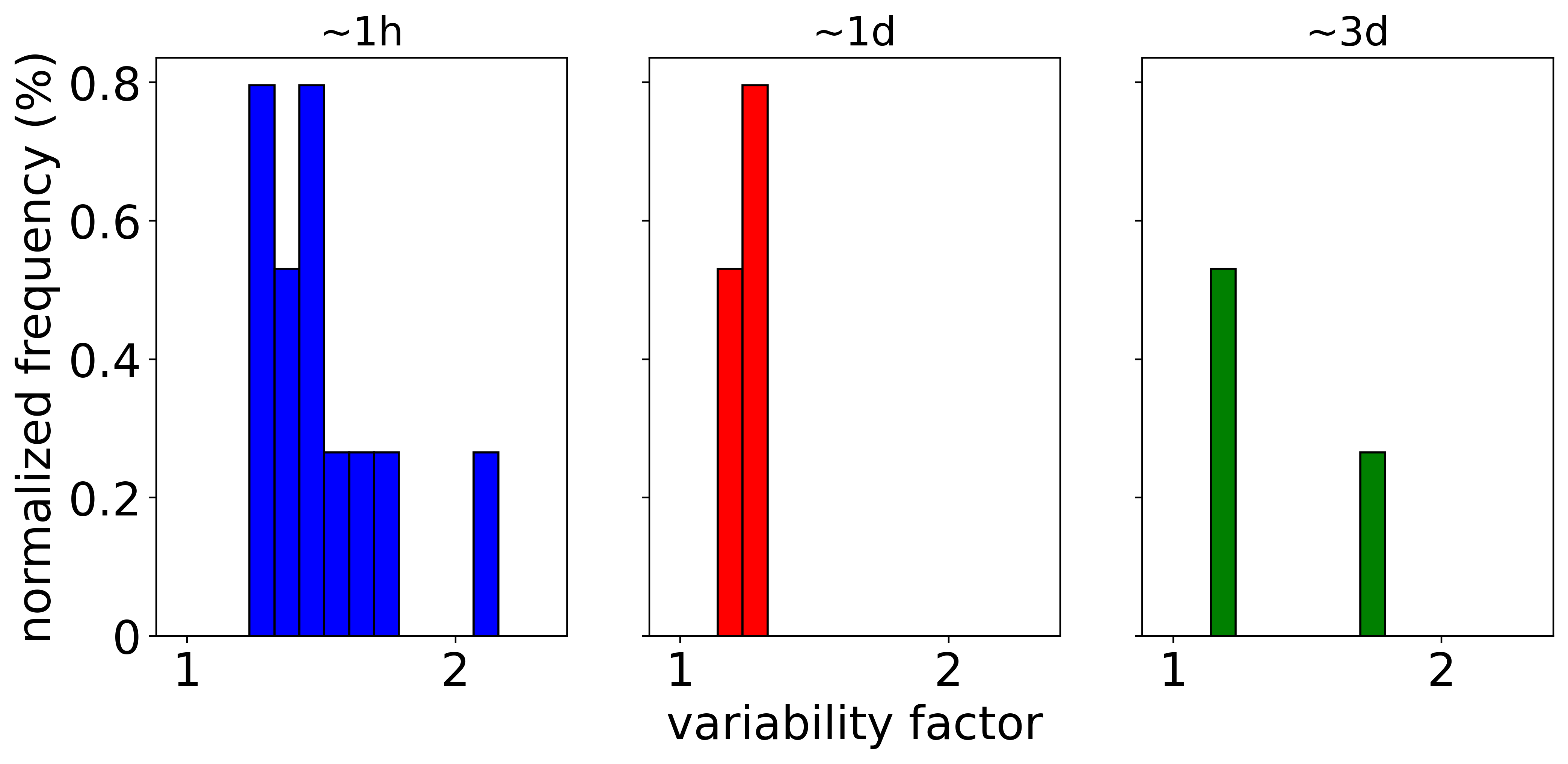}
    \caption{Normalized frequency of flux changes deemed variable in their respective timescale. The blue bars (left panel) show the variabilities over one hour, the red bars (mid panel) in one day and the green bars (right panel) over three days. The one hour timescale was sampled with all 377 sources, one day has around 260 and three days has less than 140 samples in our sample. Note that all three timescales were normalized to the total amount of detections (377), meaning that roughly 0.26~\%(1/377) on this scale translate to one source.}
    \label{fig:Var_timescales}
\end{figure}

Figure \ref{fig:Var_timescales} shows the frequencies of the variabilities by their respective timescales. 12 of the changes in flux over time that are classified as variable occur over the shortest time interval in the survey, 1 hour. 5 of the significant changes occur over a day, and the remaining three take place within three days.

The normalized frequency corresponds well with the relative occurrence of each of the timescales, with all 377 detections having been observed over one hour, 262 over one day and 144 over three days. The notable result here is that the longer timescale does not stand out contrary to expectations from literature \citep{Fischer_2023}. Given the few variable candidates outside of the 1~h timescale, no conclusion can be drawn pertaining to the variability timescales.

Recent dedicated millimeter campaigns have revealed that stellar variability operates across vastly different timescales, each tracing distinct physical mechanisms \citep{Fischer_2023}. \cite{Burton_2025} detected 463 flaring events from Proxima Cen. using ALMA, with decay timescales as short as 3–16 seconds and peak flux amplifications reaching factors of $\approx$1000, although Proxima Cen. is not a YSO. \cite{Lovell_2024} reported millimeter variability from HD 283572 resolved down to individual 30 second integrations, with a negative spectral index indicating non-thermal synchrotron emission. \cite{Miley_2025} observed a Class II disk system (J16202) showing flux decay by a factor of 8.2 over 50 minutes, again with non-thermal spectral properties. In all three cases, the rapid timescales combined with steep negative spectral indices are hallmarks of magnetic reconnection in stellar coronae or magnetospheric interactions. Note that contrary to the studies above, we do not have individual targets but populations instead, see \cite{van_Terwisga_2022}.

Our SODA survey operates in a distinctly different observational regime. While the flaring campaigns by \cite{Burton_2025, Miley_2025, Lovell_2024} probe seconds to minutes monitoring of extreme events, our survey with 2–3 epochs per source detects variability on hour to day timescales. These longer timescales may be linked to synchrotron radiation on the shorter side, while the day-long changes could also originate from accretion variability, where changes in mass transport modulate disk temperature on thermal equilibration timescales rather than on magnetic flaring timescales. Both \cite{Lovell_2024} and \cite{Miley_2025} also explore phases where accretion driven heating produces sustained flux enhancements on $\approx$hour timescales, directly bridging the non-thermal flaring regime and the accretion-driven variability.

\section{Discussion}\label{sec:Discussion}

It is typically assumed that the continuum emission from the circumstellar discs in a YSO remains relatively constant on the timescales that are probed by these observations.
With our conservative variability metric in place that yields at least 20 sources ($\approx 5.3\% $) out of all detections that can be classified as variable, we would like briefly to discuss the implications of this result in the context of previous surveys.

In comparison, our study finds more low variables (between a VF of 1.2 and $\approx$1.5) than either \cite{Vargas_Gonzalez_2023} or \cite{Johnstone_2022}, mostly due to the design of the respective surveys and their focus on higher variability in the investigation afterwards. It must be noted that the observing timescales for the JCMT transient survey in \cite{Johnstone_2022} operate on monthly cadences over 5.5 years, showing the importance of short timescale analyses again while also emphasizing the vast difference in variabilities with varying timescales. \cite{Vargas_Gonzalez_2023} focus on the clearest cases for variability, classifying all sources with a VF of >2 to be variable, where the detection is significant within a $3 s_{VF}$ range and must be above the noise envelope. This relatively high threshold is due to issues with changing in synthesized beam size. They find $\approx$6\% of sources to fit their variable criterion within the timescale of an hour, comparing well with 5.3\% in our sample although with higher thresholds and different selection criteria.

Importantly, there are also differences on a larger scale. SODA has observed clouds next to the ONC, while the JCMT transient survey was conducted over multiple star forming regions. In contrast, the survey in \cite{Vargas_Gonzalez_2023} focuses on a denser core region. The environment could be crucial in influencing the processes that drive accretion and magnetism and by extension variability. The flux measured from an individual target can vary by up to a factor 7 measured over a one hour timescale based on our study.

If our selection criteria are relaxed to a VF$>$1.15 and $\frac{VF}{s_{VF}}>4$, as many as 60 sources can be called variable candidates within the SODA sample, making at most 16~\% variable candidates out of all detections. Should this translate to other YSOs, caution must be taken when drawing conclusions from single epoch observations, as they may be variable above what is usually assumed. 

As the investigation into the used calibrators showed in the most pessimistic case of the first SB, when one observation is 3.2\% below its concatenated average due to calibrator uncertainty and the next is 3.2\% above, only 6.6\% variability\footnote{VF$=\frac{1+0.032}{1-0.032}\approx$1.066} may be attributed to calibrator fluctuations. However, even with the relaxed selection criteria, more than twice that is required to explain the variability found within this study. It should be noted that ALMA band 6 observations are sometimes assumed to have flux calibration uncertainties of up to 10\% \citep{Ansdell_2018, Yoshida_2025}.

SODA target MGM1041, classified as a lower limit, shows the most significant variability of $>$7.1 over roughly an hour as can be seen in Figure \ref{fig:dataoverview}. All lower limit sources detect the source in one or more epochs, but are not detecting it in the other(s). The SODA target MGM1041 appears in the first epoch with flux of (0.24$\pm$0.04)~mJy, whereas in the second epoch an hour later it is undetected in an image with noise level of $\approx$0.04~mJy/beam. This shows the most drastic case possible, where a source is not detected in one epoch but in another.

Assuming that the variability found in the SODA sample is not unique to the observed sources, similar variability can be expected in any given sample of unresolved or low-resolution observations of Class II YSOs when observed at millimeter wavelengths and possibly others. At the same time, we expect this type of variability to only affect the central beam of high-resolution observations.

It should be noted here that \cite{Vargas_Gonzalez_2021} and \cite{Vargas_Gonzalez_2023} conducted studies of sources solely in the more dense ONC core and were targeting minute-scale flares where we looked at a large, spread out area adjacent to the ONC on longer timescales. Further investigation in other star-forming regions are required to test this assumption and to classify any possible variability, to look for systematic differences between individual regions. 

We would like to stress that extensive surveys would be impacted most from the variability if it is not taken into account. Individual disk masses in large surveys could be systematically biased by as much as a factor of 7 in the worst case based solely on our investigation, and possibly more as evident in literature \citep{Salter_2008, Mairs_2019, Lovell_2024}.The need to analyze variability in large surveys is thus plainly clear and follows from this investigation.

Akin to e.g. \cite{van_Terwisga_2020}, \cite{Vargas_Gonzalez_2023} and \cite{Marquez_2024}, we cannot constrain the exact physical mechanism creating the observed variability in our sample. The observed timescales in the range of hours suggest (gyro-)synchrotron radiation for the shortest changes, as for example laid out in \cite{Benz_2010}. The short time span and the relatively high amplitude make this the most likely origin. The longer timescales could also stem from thermal origins. It is also possible that only a part of a variability timescale was measured in the data herein, which could explain some of the lower variables that were only measured over short times.

No obvious connection to binarity was found in the variable candidates. This may well be due to only 65 multiple systems having been found in \cite{Ricciardi_2025} and only 20 sources being classified as variable candidates in this study. For a deeper look into binarity in the SODA sample see \cite{Ricciardi_2025}, who combined SODA data with ESO-VISTA data to investigate them for multiplicity. Stellar masses of most of the sources from [MGM2012] are known with no clear correlation between stellar mass and variability or any other metric. It should be noted that not all of the stellar masses of our variable candidates seem to be known \citep{DaRio_2016}. In a comparison with WISE variability, the sources that are covered by WISE show no correlation with the variability of this work.

\section{Conclusion}\label{sec:Conclusion}
This study reveals that short term millimeter wavelength variability in unresolved YSO Class II is more common than previously assumed.
\begin{itemize}
    \item We report our findings of variability over short time scales ranging from an hour to a few days. These changes in flux occur at the 2~mm wavelength in band 6 of ALMA in observations carried out of the L1641 and L1647 molecular clouds near Orion. All sources had observation timescales of an hour, while $\approx$260 have been observed with a timescale of one day, and $\approx$140 over three days, respectively.
    \item From a sample of all detections in the SODA survey \citep[see][]{van_Terwisga_2022} 5.3\% show statistically significant flux changes, with one source reaching a factor of >7 in just one hour.
    \item The variable candidates have been selected based on their normalized flux variation compared to the whole sample and the variability factor within the individual observations of one concatenated source. We count a source as a variable candidate if its VF is above 1.2 and the S/N of the VF is 7 or higher, yielding 20 variable candidates. If criteria are relaxed to VF > 1.15 and the S/N to 4, we find 60 sources ($\approx$16\%) to be variable candidates.
    \item We use two independent metrics for variability, the variability factor VF and the standard deviation of the flux ($\sigma_F$) relative to the flux of the concatenated source $\frac{\sigma_F}{F_0}$. We compare the results of one metric with the other to assess variability. 
    \item If variability is as common as suggested by this work, some observations in the millimeter range that assume no significant variability over short times may be worth revisiting. Even single observations (one EB) could carry non-negligible uncertainties, given that 16\% of all detections show variability >1.15. Conclusions drawn from large surveys in particular may be prone to a systematic bias.
    \item The variability encountered in the SODA sample may be explained by magnetospheric accretion for the larger timescales of one to three days. The shorter changes, particularly variability over an hour exceeding a VF of 2, is stronger than what is usually attributed to accretion. Instead, we direct to the studies \cite{Bower_2003},\cite{van_Terwisga_2020}, \cite{Vargas_Gonzalez_2023} and \cite{Marquez_2024} that suggest (gyro-)synchrotron radiation due to magnetic reconnection from inside the YSO and not to the dust itself as the mechanism behind the variability.
    \item The relatively low calibrator variability of just 3.2$\%$ at most cannot be held accountable for the changes in flux, see Figure \ref{fig:phasecal} and Figure \ref{fig:VarSN} for details. Even with 10\% assumed calibrator variability, multiple sources stand out as significantly more variable than explainable with calibration.
\end{itemize}

Further studies on the topic in a variety of star-forming regions are necessary to assess millimeter variability with our methods more thoroughly. Variability exceeding calibrator uncertainties and not taken into account on short timescales poses a problem in form of a possible source of bias for a mass of past radio observations and must be quantified to allow serious observations without large, unknown uncertainties.

\begin{acknowledgements}

D. Krebs thanks the Institute for Astrophysics of the University of Vienna and the Centre for Astrophysics Research of the University of Hertfordshire for their support.
This project has received funding from the European Research Council (ERC) under the European Union’s Horizon 2020 research and innovation programme (Grant agreement No. 851435).
The computational results have been achieved using the Austrian Scientific Computing (ASC) infrastructure.
This paper makes use of the following ALMA data ADS/JAO.ALMA\#2019.1.01813.S. ALMA is a partnership of ESO (representing its member states), NSF (USA) and NINS (Japan), together with NRC (Canada), MOST and ASIAA (Taiwan), and KASI (Republic of Korea), in cooperation with the Republic of Chile. The Joint ALMA Observatory is operated by ESO, AUI/NRAO and NAOJ.

\end{acknowledgements}

\bibliography{mybib}
\end{document}